\documentclass[aps,prb,twocolumn,showpacs,preprintnumbers,amsmath,amssymb,superscriptaddress]{revtex4}%

\usepackage{graphicx}%
\usepackage{dcolumn}
\usepackage{amsmath}
\usepackage{color}

\makeatletter
\def\btt#1{\texttt{\@backslashchar#1}}%
\DeclareRobustCommand\bblash{\btt{\@backslashchar}}%
\makeatother

\begin{document}


\title{Muon-spin rotation measurements of the penetration depth in Li$_2$Pd$_3$B}

\author{R.~Khasanov}
\affiliation{Physik-Institut der Universit\"{a}t Z\"{u}rich,
Winterthurerstrasse 190, CH-8057, Z\"urich, Switzerland}
\affiliation{Laboratory for Muon Spin Spectroscopy, PSI, CH-5232
Villigen PSI, Switzerland}
%
\author{I.L.~Landau}
\affiliation{Kapitza Institute for Physical problems, 117334 Moscow,
Russia}
\author{C.~Baines}
\affiliation{Laboratory for Muon Spin Spectroscopy, PSI, CH-5232
Villigen PSI, Switzerland}
\author{F.~La~Mattina}
\affiliation{Physik-Institut der Universit\"{a}t Z\"{u}rich,
Winterthurerstrasse 190, CH-8057, Z\"urich, Switzerland}
\author{A.~Maisuradze}
\affiliation{Physik-Institut der Universit\"{a}t Z\"{u}rich,
Winterthurerstrasse 190, CH-8057, Z\"urich, Switzerland}
\author{K.~Togano}
\affiliation{National Institute for Materials Science
  1-2-1, Sengen, Tsukuba, Ibaraki 305-0047, Japan}
\author{H.~Keller}
\affiliation{Physik-Institut der Universit\"{a}t Z\"{u}rich,
Winterthurerstrasse 190, CH-8057, Z\"urich, Switzerland}

\begin{abstract}
Measurements of the magnetic field penetration depth $\lambda$ in
the ternary boride superconductor Li$_2$Pd$_3$B ($T_c\simeq7.3$~K)
have been carried out by means of muon-spin rotation ($\mu$SR). The
absolute values of $\lambda$, the Ginzburg-Landau parameter
$\kappa$, and the first $H_{c1}$ and the second $H_{c2}$ critical
fields at $T=0$ obtained from $\mu$SR were found to  be
$\lambda(0)=252(2)$~nm, $\kappa(0)=27(1)$,
$\mu_0H_{c1}(0)=9.5(1)$~mT, and $\mu_0H_{c2}(0)=3.66(8)$~T,
respectively. The zero-temperature value of the superconducting gap
$\Delta_0=$1.31(3)~meV was found, corresponding to the ratio
$2\Delta_0/k_BT_c=4.0(1)$. At low temperatures $\lambda(T)$
saturates and becomes constant below $T\simeq 0.2T_c$, in agreement
with what is expected for s-wave BCS superconductors. Our results
suggest that Li$_2$Pd$_3$B is a s-wave BCS superconductor with the
only one isotropic energy gap.
\end{abstract}

\maketitle

\section{Introduction}

The discovery of superconductivity in the ternary boride
superconductors Li$_2$Pd$_3$B and Li$_2$Pt$_3$B has attracted
considerable interest in the study of these
materials.\cite{Togano04,Badica04,Yokoya05,Nishiyama05,Yuan05,Takeya05}
It is believed now that superconductivity in both above mentioned
compounds is most likely mediated by phonons. It stems from
photoemission,\cite{Yokoya05} nuclear magnetic resonance (NMR),
\cite{Nishiyama05} and specific heat experiments. \cite{Takeya05}
Moreover, the observation of a Hebel-Slichter peak in the $^{11}$B
spin-lattice relaxation rate in Li$_2$Pd$_3$B strongly supports
singlet pairing.\cite{Nishiyama05} However, experimental results
concerning the structure of the superconducting energy gap are still
controversal. On the one hand, NMR data of Li$_2$Pd$_3$B (Ref.
\onlinecite{Nishiyama05}) and specific heat data of Li$_2$Pd$_3$B
and Li$_2$Pt$_3$B (Ref. \onlinecite{Takeya05}) can be well explained
assuming conventional superconductivity. On the other hand, recent
measurements of the magnetic field penetration depth $\lambda$
suggest unconventional behavior of both compounds, namely double-gap
superconductivity in Li$_3$Pd$_2$B and nodes in the energy gap in
Li$_3$Pt$_2$B.\cite{Yuan05,Yuan05a} This contradiction is serious
and shows that further experimental investigations of these
compounds are needed.

In this paper, we report a systematic study of magnetic field
penetration depth $\lambda$ in Li$_2$Pd$_3$B by means of
transverse-field muon-spin rotation (TF-$\mu$SR) (the detailed
description of TF-$\mu$SR technique in connection with $\lambda$
studies can be found, e.g., in Ref.~\onlinecite{Zimmermann95}).
Measurements were performed down to 30~mK in a series of fields
ranging from 0.02~T to 2.3~T. For all magnetic fields studied
(0.02~T, 0.1~T, 0.5~T, 1~T, and 2.3~T) no sign of a second
superconducting gap was detected. All our results may be well
explained by assuming conventional superconductivity with the only
{\it one isotropic} energy gap. The absolute values of $\lambda$,
the Ginzburg-Landau parameter $\kappa$, and the first ($H_{c1}$) and
the second ($H_{c2}$) critical fields at $T=0$ obtained from $\mu$SR
were found to be $\lambda(0)=252(2)$~nm, $\kappa(0)=27(1)$,
$\mu_0H_{c1}(0)=9.5(1)$~mT, and $\mu_0H_{c2}(0)=3.66(8)$~T,
respectively. The zero temperature value of the superconducting gap
$\Delta_0$ was found to be 1.31(3)~meV that corresponds to the ratio
$2\Delta_0/k_BT_c=4.0(1)$.

The paper is organized as follows: In Sec.~\ref{sec:experimental} we
describe the sample preparation procedure and the TF-$\mu$SR
technique in connection with $\lambda(T)$ measurements. In
Sec.~\ref{subsec:Hc1-Hc2} we discuss the temperature dependence of
the second critical field $H_{c2}$. In Sec.~\ref{subsec:lambda_vs_H}
we present the calculation of the absolute value of $\lambda$ and
the magnetic field dependence of the second moment of the $\mu$SR
line. Sec.~\ref{subsec:lambda_vs_T} comprises studies of the
temperature dependence of $\lambda$. The conclusions follow in
Sec.~\ref{sec:conclusion}.

\section{Experimental details}\label{sec:experimental}

The Li$_2$Pd$_3$B polycrystalline sample was prepared by two-step
arc-melting.\cite{Togano04} First, a binary Pd$_3$B alloy was
prepared by conventional arc-melting from the mixture of Pd(99.9\%)
and B(99.5\%).  The alloying of Li was done in the second
arc-melting, in which a small piece of Pd$_3$B alloy was placed on a
Li($>99$\%) plate. Once the Pd$_3$B alloy melted, the reaction with
Li occurred and developed very fast, forming a small button specimen
(around 300~mg). Since the loss of Li was inevitable, the Li
concentration in the final sample was estimated from the weight
change.  The deviation of the Li concentration from the
stoichiometric one was less than 1\% for the specimens used in this
experiment.
\begin{figure}[htb]
\includegraphics[width=0.9\linewidth]{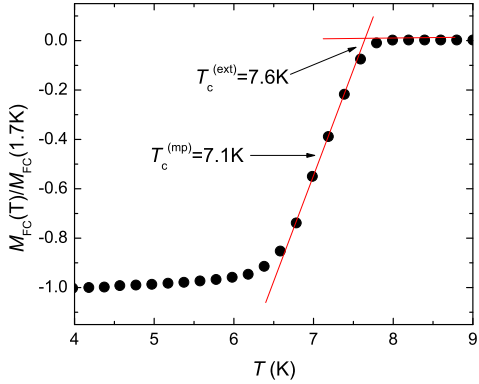}
%
\caption{(Color online) Field-cooled magnetization curve normalized
to their value at 1.7~K [$M_{FC}(T)/M_{FC}(1.7~K)$] for $\mu_0H =
0.5$ mT. The solid line is the best linear fit to the steepest part
of the $M_{FC}(T)$ curve. $T_c^{(mp)}$ and $T_c^{(ext)}$ are
indicated in the figure and discussed in the text.}
 \label{fig:Fig1}
\end{figure}

Field--cooled magnetization ($M_{FC}$) measurements were performed
with a SQUID magnetometer in fields ranging from 0.5~mT to 4~T and
temperatures between 1.75~K and 10~K. The $M_{FC}(T)$ curve for
$\mu_0H= 0.5$~mT is shown in Fig.~\ref{fig:Fig1}. The
superconducting transition is rather broad indicating that the
sample is not particularly uniform, i.e., the superconducting
transition temperature may be evaluated only approximately. The
middle-point of the transition corresponds to $T_c^{(mp)} =  7.1$ K,
while the linear extrapolation of the steepest part of the $M(T)$
curve to $M=0$ results in $T_c^{(ext)} = 7.6$ K (see Fig.
\ref{fig:Fig1}).

TF-$\mu$SR experiments were performed at the $\pi$M3 beam line at
Paul Scherrer Institute (Villigen, Switzerland). The Li$_2$Pd$_3$B
sample was field cooled from above $T_c$ down to 30~mK in fields of
2.3~T, 1~T and 0.5~T, and down to $\simeq$1.6~K in fields of 0.1~T
and 0.02~T.
\begin{figure}[htb]
 \includegraphics[width=1.0\linewidth]{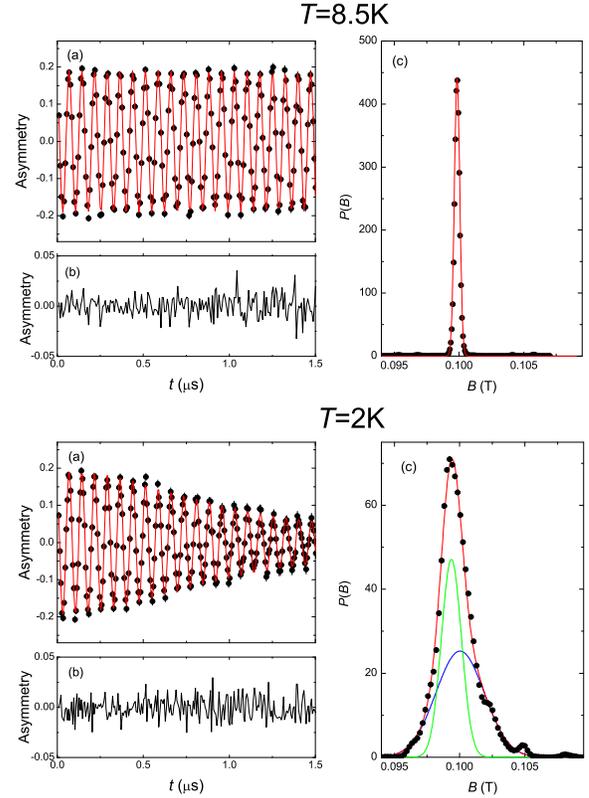}
\caption{(Color online) $\mu$SR time spectra and magnetic field
distributions of Li$_2$Pd$_3$B taken above ($T=8.5$~K) and below
($T=2$~K) the superconducting transition temperature $T_c$ in
$\mu_0H = 0.1$~T: (a) the muon-time spectra, (b) difference between
the one Gaussian (upper panle) and two Gaussian (lower panel) fits
and experimental data, (c) internal field distributions inside the
Li$_2$Pd$_3$B sample. The lines represent the best fit with the
Gaussian line-shapes. See text for details.}
 \label{fig:Fourier}
\end{figure}
In the transverse-field geometry the local magnetic field
distribution $P(B)$ inside the superconducting sample in the mixed
state, probed by means of the TF-$\mu$SR technique, is determined by
the values of the coherence length $\xi$ and the magnetic field
penetration depth $\lambda$. In extreme type-II superconductors
($\lambda\gg\xi$) the $P(B)$ distribution is almost independent on
$\xi$ and the second moment of $P(B)$ line becomes simply
proportional to $1/\lambda^4$.\cite{Brandt88,Brandt03}

The $\mu$SR signal was recorded in the usual time-differential way
by counting positrons from decaying muons as a function of time. The
time dependence of the positron rate is given by the expression
\cite{msr}
\begin{equation}
 {N(t)} = N_0 {1\over\tau_\mu} e^{-t/\tau_\mu}
  \left[ 1 + a P(t) \right] + bg \; ,
\label{eq:N_t}
\end{equation}
where $N_0$ is the normalization constant, $bg$ denotes the
time-independent background, $\tau_\mu = 2.19703(4) \times
10^{-6}$~s is the muon lifetime, $a$ is the maximum decay
asymmetry for the particular detector telescope ($a\sim 0.18$ in
our case), and $P(t)$ is the polarization of the muon ensemble:
\begin{equation}
P(t)=\int P(B)\cos(\gamma_{\mu}Bt+\phi)dB \; .
 \label{eq:P_t}
\end{equation}
Here $\gamma_{\mu} = 2\pi\times135.5342$~MHz/T is the muon
gyromagnetic ratio and $\phi$ is the angle between the initial muon
polarization and the effective symmetry axis of a positron detector.
$P(t)$ can be linked to the internal field distribution $P(B)$ by
using the algorithm of Fourier transform.\cite{msr}

The $P(t)$ and $P(B)$ distributions inside the Li$_2$Pd$_3$B sample
in the normal ($T>T_c$) and in the mixed state ($T<T_c$) after field
cooling in a magnetic field of 0.1~T are shown in
Fig.~\ref{fig:Fourier}. The $P(B)$ distributions were obtained from
the measured $P(t)$ by using the fast Fourier transform procedure
based on the maximum entropy algorithm.\cite{Rainford94} In the
normal state, a symmetric line at the position of the external
magnetic field with a broadening arising from the nuclear magnetic
moments is seen. Below $T_c$ the field distribution is broadened and
asymmetric. In order to account for the asymmetric field
distribution, $\mu$SR time spectra obtained below $T_c$ were fitted
by two Gaussian lines:\cite{Serventi04,Khasanov05}
\begin{eqnarray}
P(t)=\sum_{i=1}^2A_i \exp(-\sigma_i^2t^2/2) \cos(\gamma_{\mu}B_i
t+\phi) \;,
\label{eq:gauss}
\end{eqnarray}
where $A_i$, $\sigma_i$, and $B_i$ are the asymmetry, the Gaussian
relaxation rate, and the first moment of the $i$-th line,
respectively. At $T>T_c$, the analysis is simplified to a single
line with $\sigma_{nm} \sim 0.1$~MHz arising from the nuclear
moments of the sample. Eq.~(\ref{eq:gauss}) is equivalent to the
field distribution \cite{Khasanov05}
\begin{equation}
P(B)=\gamma_{\mu}\sum_{i=1}^2{A_i \over \sigma_i}
\exp\left(-{\gamma_{\mu}^2(B-B_i)^2 \over 2\sigma_i^2}\right) \; .
\label{eq:P_B}
\end{equation}
The solid lines in Figs.~\ref{fig:Fourier}~(a) represent the best
fit with the one (upper panel) and two Gaussian lines (lower panel)
to the $\mu$SR time spectra. The corresponding $P(B)$ lines are
shown in Figs.~\ref{fig:Fourier}~(c).
For this distribution the mean field and the second moment are
\cite{Khasanov05}
\begin{equation}
\langle B \rangle=\sum_{i=1}^2{A_i B_i \over A_1+A_2} \;
\label{eq:B_mean}
\end{equation}
and
\begin{equation}
\langle \Delta B^2
\rangle=\frac{\sigma^2}{\gamma^2_\mu}=\sum_{i=1}^2{A_i \over
A_1+A_2} \left[ \frac{\sigma_i^2}{\gamma_{\mu}^2} +[B_i- \langle B
\rangle]^2 \right] \;.
\label{eq:dB}
\end{equation}

The superconducting part of the square root of the second moment
$\sigma_{sc}$ was then obtained by subtracting the contribution of
nuclear moments $\sigma_{nm}$ measured at $T>T_c$ as
$\sigma_{sc}^2=\sigma^2 - \sigma_{nm}^2$. From the known value of
$\sigma_{sc}$ the absolute value of $\lambda$ can be evaluated using
the following relation
\begin{eqnarray}
 \sigma_{sc}[\mu {\rm s}^{-1}]=4.83\times10^4
(1 - H/H_{c2}) \nonumber \\
 \left[1 + 1.21\left(1 - \sqrt{H/H_{c2}}\right)^3\right]&
\lambda^{-2}[{\rm nm}] \; ,
 \label{eq:sigma_vs_h}
\end{eqnarray}
which describes the field variation of  $\sigma_{sc}$ for an ideal
triangular vortex lattice.\cite{Brandt03} Note, that according to
Ref.~\onlinecite{Brandt03}, Eq.~(\ref{eq:sigma_vs_h}) does not hold
for very low magnetic inductions .

\section{Experimental Results and Discussion}
 \label{seq:results_and_discussions}

\subsection{Temperature dependence of the upper critical field}
 \label{subsec:Hc1-Hc2}

 \begin{figure}[htb]
\includegraphics[width=1.0\linewidth]{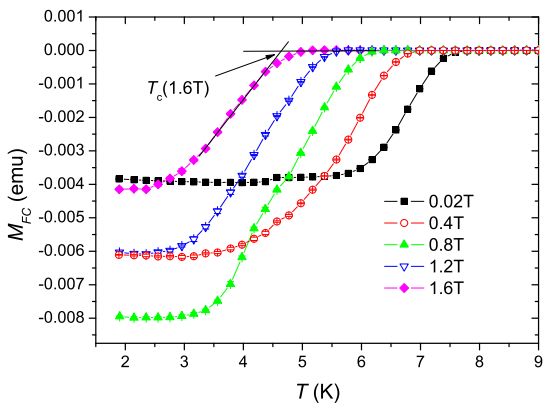}
%
\caption{(Color online) Field-cooled magnetization curves $M_{FC}(T)$ of
Li$_2$Pd$_3$B taken at various magnetic fields: from the left to the
right $\mu_0H=$ 1.6~T, 1.2~T, 0.8~T, 0.4~T, and 0.02~T. The
paramagnetic background was subtracted.}
 \label{fig:magnetization}
\end{figure}

\begin{figure}[htb]
\includegraphics[width=1.0\linewidth]{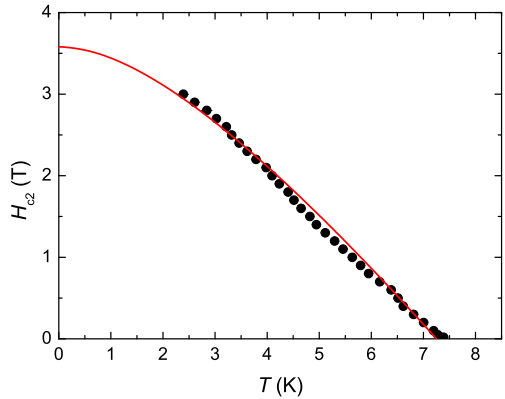}
%
\caption{(Color online) Temperature dependence of the upper critical
field $H_{c2}$ of Li$_2$Pd$_3$B. The solid line is the fit using the
WHH model. The fit parameters are listed in the text.}
 \label{fig:Hc2}
\end{figure}

The field-cooled magnetization $M_{FC}(T)$ curves for several
magnetic fields $H$ are shown in Fig.~\ref{fig:magnetization}. The
transition temperature $T_c$ was taken from the linearly
extrapolated $M(T)$ curves in the vicinity of $T_c$ with $M=0$ line.
For each particular field $H$ the corresponding transition
temperature $T_c(H)$ was taken as the temperature where
$H=H_{c2}(T=T_c)$.
The resulting $H_{c2}(T)$ dependence is presented in
Fig.~\ref{fig:Hc2}.
%
It is seen that $H_{c2}(T)$ data can be satisfactory fitted with the
model provided by the Werthamer-Helfand-Hohenberg (WHH)
theory.\cite{Werthamer66} The fit yields $H_{c2}(0) = 3.58(10)$~T
and $T_c = 7.25$ K. Note, that the value of the transition
temperature obtained from the fit lies between $T_c^{(mp)}$ and
$T_c^{(ext)}$ introduced in Fig.~\ref{fig:Fig1}, i.e., it is in
agreement with low-field magnetization measurements. In the
following, the $H_{c2}(T)$ curve presented in Fig.~\ref{fig:Hc2} is
used to analyze the $\mu$SR data (see
Sec.~\ref{subsec:lambda_vs_T}).

To summarize, the temperature dependence of the second critical
field can be well described within the WHH theory. The absolute
value of the second critical field at $T=0$ was found to be
3.58(10)~T in agreement with results of Badica {\it et al.}
Ref.~\onlinecite{Badica04}.

\subsection{Magnetic field dependence of the second moment of $\mu$SR line}
 \label{subsec:lambda_vs_H}

\begin{figure}[htb]
\includegraphics[width=1.0\linewidth]{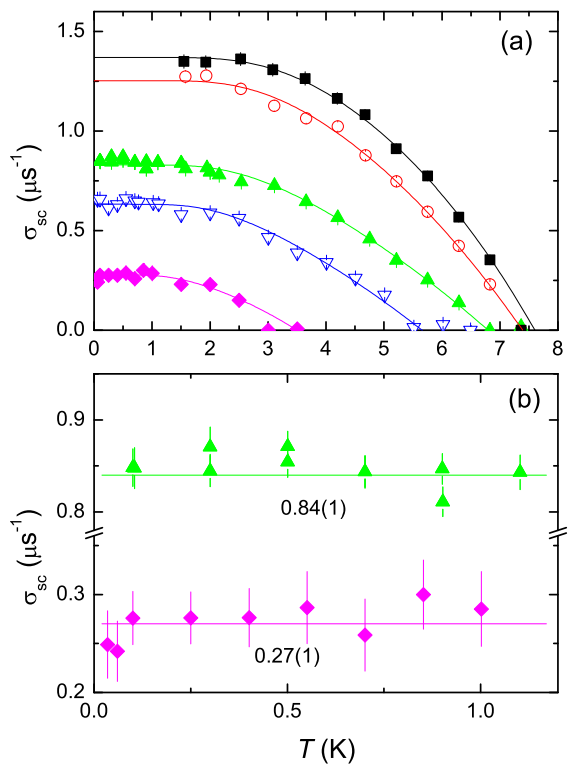}
%
\caption{(Color online) (a) Temperature dependencies of
$\sigma_{sc}$ measured in various magnetic fields (from top to the
bottom) 0.02~T, 0.1~T, 0.5~T, 1~T, and 2.3~T. The solid lines are
guides to the eye. (b) Low-temperature part of $\sigma_{sc}(T)$ for
$H=$0.5~T and 2.3~T. }
 \label{fig:lambda_vs_T}
\end{figure}

\begin{figure}[htb]
\includegraphics[width=1.0\linewidth]{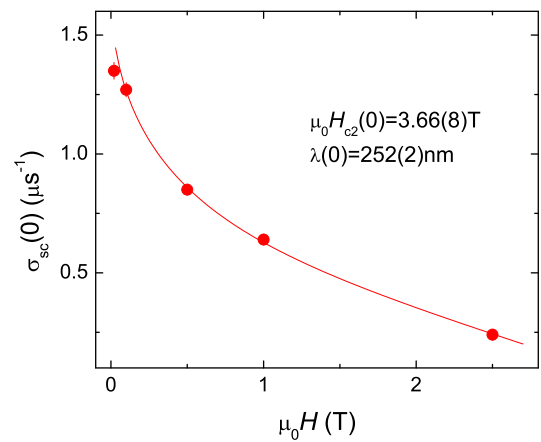}
\caption{(Color online) Field dependence of the zero-temperature
$\mu$SR depolarization rate $\sigma_{sc}(0)$ in Li$_2$Pd$_3$B
sample. The solid line corresponds to a fit of
Eq.~(\ref{eq:sigma_vs_h}) to the experimental data with the
parameters given in the text. }
 \label{fig:lambda_vs_H_muSR}
\end{figure}

In Fig.~\ref{fig:lambda_vs_T}~(a) the temperature dependences  of
$\sigma_{sc}$ for $\mu_0 H=0.02$~T, 0.1~T, 0.5~T, 1~T, and 2.3~T are
shown.  For $\mu_0H=$0.5, 1~T, and 2.3~T, $\sigma_{sc}(T)$ was
measured down to 30~mK. It is seen, that below 1.5~K $\sigma_{sc}$
is practically temperature independent
[Fig.~\ref{fig:lambda_vs_T}(b)]. Bearing in mind, that the
temperature dependence of $\sigma_{sc}$ saturates at low
temperatures, the values of $\sigma_{sc}(T=0)$ can be reliably
evaluated  for all magnetic fields. The results are plotted in
Fig.~\ref{fig:lambda_vs_H_muSR} as a function of $H$. For
$\mu_0H=0.5$~T, 1.0~T, and 2.3~T $\sigma_{sc}(0)$ was obtained from
the zero slope linear fit of the $T\leq1.5$~K data [see
Fig.~\ref{fig:lambda_vs_T}(b)] and for $\mu_0H=0.02$~T and 0.1~T
$\sigma_{sc}(0)$ was assumed to be equal to $\sigma_{sc}$ at the
lowest measured temperature ($T\simeq1.55$~K). The solid line in
Fig.~\ref{fig:lambda_vs_H_muSR} represents the result of the fit of
Eq.~(\ref{eq:sigma_vs_h}) to the experimental data with $H_{c2}(0) =
3.66(8)$~T and $\lambda(0) = 252(2)$~nm.
%
As it was already mentioned above, Eq.~(\ref{eq:sigma_vs_h}) does
not hold for the very low magnetic field. For this reason  the data
point for $\mu_0H=0.02$~T was excluded from the fit. It is seen
(Fig.~\ref{fig:lambda_vs_H_muSR}), that the theoretical curve
perfectly matches all data points for $\mu_0H \ge 0.1$ T. The value
of $\sigma_{sc}$ for $\mu_0H = 0.02$~T lies slightly below the
theoretical curve, as expected. It is important to emphasize that
the value of $\mu_0H_{c2}(0) = 3.66(8)$~T obtained from the fit of
$\sigma_{sc}(0,H)$ data coincides within the error with
$\mu_0H_{c2}(0) = 3.58(10)$~T, evaluated in the previous section
from the directly measured $H_{c2}(T)$. This good agreement between
the values of $H_{c2}$, obtained from the two completely different
experiments, clearly demonstrates the validity of our analysis.

The above presented experiments clearly demonstrate, that $\lambda$,
evaluated from $\mu$SR measurements, is indeed {\it magnetic field
independent}, as one would expect in case of a conventional
superconductor with isotropic energy gap.\cite{Kadono04} On the
other hand, in superconductors with nodes in the gap and isotropic
double-gap superconductors like MgB$_2$, $\lambda$, evaluated in the
same way, increases with increasing magnetic field (see e.g.,
Refs.~\onlinecite{Serventi04}, \onlinecite{Kadono04}, and
\onlinecite{Sonier00}). Thus, the fact, that the $\sigma_{sc}(0)$
versus $H$ dependence is perfectly described by the field
independent $\lambda$ (see Fig.~\ref{fig:lambda_vs_H_muSR}), implies
that Li$_2$Pd$_3$B is a {\it conventional single-gap}
superconductor.

The zero--temperature value of the superconducting coherence  length
$\xi(0)$ may be estimated from  $H_{c2}(0)$ as $\xi(0)=(\Phi_0/2\pi
H_{c2}(0))^{0.5}$, which results in $\xi(0) = 9.5(2)$~nm ($\Phi_0$
is the magnetic flux quantum). Using the values of $\lambda(0)$ and
$\xi(0)$, one can also evaluate the zero-temperature value of the
Ginzburg-Landau parameter $\kappa(0) = \lambda(0)/\xi(0) \approx
27(1)$. The value of the first critical field can also be calculated
by means of Eq.~(4) from Ref.~\onlinecite{Brandt03} as $\mu_0H_{c1}
= 9.5(1)$ mT. It is remarkable that all superconducting
characteristics of Li$_2$Pd$_3$B could be obtained solely from
$\mu$SR experiments.

To summarize, the magnetic field dependence of the superconducting
part of the $\mu$SR depolarization rate $\sigma_{sc}$ is well
described within the Ginzburg-Landau theory for anisotropic
single-gap superconductors. The zero-temperature values of the first
and the second critical fields, the magnetic penetration depth, the
coherence length, and the Ginzburg-Landau parameter were found to be
$\mu_0H_{c2}(0)=3.66(8)$~T, $\mu_0H_{c1}=9.5(1)$~mT,
$\lambda(0)=252(2)$~nm, $\xi(0)=9.5(2)$~nm, and $\kappa=27(1)$,
respectively.

\subsection{Temperature dependence of $\lambda$}
 \label{subsec:lambda_vs_T}

\begin{figure}[htb]
\includegraphics[width=1.0\linewidth]{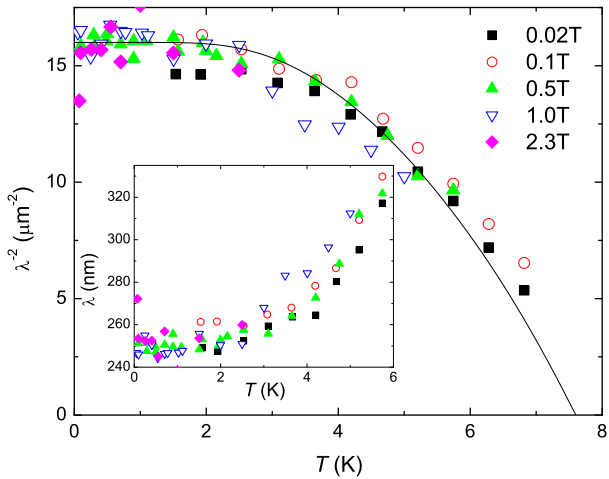}
\caption{(Color online) $1/\lambda^2$ versus temperature. The solid
line represents the theoretical curve provided by the BCS theory for
$2\Delta_0 = 4k_BT_c$ with $T_c = T_c^{ext} = 7.6$ K (see
Fig.~\ref{fig:Fig1}). The inset shows $\lambda(T)$. The errors are
not shown for clarity. }
 \label{fig:lambda-T}
\end{figure}

Eq.~(\ref{eq:sigma_vs_h}) implies that $\sigma_{sc}$ depends on
$\lambda$ and the reduced magnetic field $H/H_{c2}$. It is also
clear that $\sigma_{sc}$ vanishes for $H \geq H_{c2}(T)$. This
means, that in order to obtain the temperature dependence of
$\lambda$ from the $\sigma_{sc}(T)$ curves, the temperature
dependence of $H/H_{c2}$ must be taken into account. In our
calculations, the $H_{c2}(T)$ curve provided by the solid line in
Fig.~\ref{fig:Hc2} was used. Due to the known value of $\kappa =
27(1)$, we are not limited by Eq.~(\ref{eq:sigma_vs_h}) but can
directly apply the numerical calculations of Brandt.\cite{Brandt03}
In this case, the results collected in the lowest magnetic field
($\mu_0H=0.02$~T) can also be used.
The resulting temperature dependence of $\lambda$ is shown in the
Fig.~\ref{fig:lambda-T}. Remarkably, all $\lambda$ points,
reconstructed from $\sigma_{sc}$ measured in various magnetic
fields, collapse onto a single $\lambda(T)$ curve. One should
emphasize, that in this reconstruction no adjustable parameter was
used. The $\sigma_{sc}(T)$ dependences were obtained by means of
$\mu$SR, while the $H_{c2}(T)$ curve was measured in a completely
different set of magnetization experiments (see
Sec.~\ref{subsec:Hc1-Hc2}).

Considering the temperature dependence of $\lambda$, one should
note, that it can be better described by the BCS theory for a
moderate coupling with the zero-temperature energy gap $2\Delta_0 =
4.0(1)k_B T_c$ [$\Delta(0)=1.31(3)$~meV] rather than for the weak
coupling limit. This is in agreement with the value $2\Delta_0 =
3.94k_BT_c$ obtained recently from specific-heat
measurements.\cite{Takeya05}
The upward deviations of the experimental data-points from  the
theoretical curve at higher temperatures (see Fig.
\ref{fig:lambda-T}) is expected for nonuniform samples (e.g. for the
samples where the transition temperatures are distributed via a
certain range $\Delta T_c$). At temperatures close to $T_c$
$\lambda^{-2} \sim (T_c - T)$. Assuming that the critical
temperature varies throughout  the sample from $T_{c0} - \Delta T_c$
to $T_{c0}$, then the relative variation of $\lambda^{-2}$ for
$T<T_{c0}-\Delta T_c$ is proportional to $[1 - \Delta T_c/(T_{c0} -
T)]^{-1}$, i.e., it increases with increasing temperature. The value
of $\sigma$, resulting from $\mu$SR experiments in the case of
nonuniform samples, may be written as $\sigma^2 = 1/V_0\int
\sigma_{local}^2dV$ ($\sigma_{local}$ is the local value of $\sigma$
in the unit volume $dV$ and $V_0$ is the sample volume). Taking into
account that $\sigma_{local}^2 \sim 1/\lambda_{local}^4$ this
averaging implies that the parts of the sample with the smallest
values of $\lambda$ (the highest values of $T_c$) provide the main
contribution to $\sigma_{sc}$. At lower temperatures, however,
$\lambda$ become more uniform throughout the sample and
$\sigma_{sc}$ can be used for reliable calculations of $\lambda(T)$.

Recently, Yuan {\it et al.} studied the magnetic field  penetration
depth in Li$_2$Pd$_3$B by means of self-inductance
technique.\cite{Yuan05,Yuan05a} It was obtained that a considerable
increase of $\lambda$ starts already at temperatures well below 1 K.
Since such a behavior contradicts conventional BCS theory, the
authors assumed  the presence of the second superconducting gap. Our
results (see Fig.~\ref{fig:lambda-T}) are quite different.
$\lambda(T)$, evaluated from $\mu$SR measurements, is practically
temperature independent below 2~K in complete agreement with
conventional single-gap theories of superconductivity.
We argue that the most probable reason for the above mentioned
disagreement in the $\lambda(T)$ dependence comes from the
difference in the experimental techniques. The self-inductance
technique used in Refs.~\onlinecite{Yuan05} and \onlinecite{Yuan05a}
provides information about the properties of the surface of the
sample. It is likely that in this Li-containing compound the
properties of the surface layer of the sample are different from
those of the bulk.

To summarize, in the whole  temperature range (from $T_c$ down to
30~mK) the temperature dependence of $\lambda$ is consistent with
what is expected for a single-gap s--wave BCS superconductor. The
value of the superconducting gap was found to be
$\Delta_0=1.31(3)$~meV, that corresponds to the ratio
$2\Delta_0/k_BT_c=4.0(1)$.

\section{Conclusions}\label{sec:conclusion}

Muon-spin rotation and magnetization studies were performed on the
ternary boride superconductor Li$_2$Pd$_3$B ($T_c\simeq7.3$~K). The
main results are: (i) The absolute values of $\lambda$, $\xi$,
$\kappa$, $H_{c1}$, and $H_{c2}$ at zero temperature obtained from
$\mu$SR are: $\lambda(0)=252(2)$~nm, $\xi(0) = 9.5$ nm, $\kappa(0) =
27(1)$, $\mu_0H_{c1}(0)=9.5(1)$~mT, and $\mu_0H_{c2}(0) =
3.66(8)$~T. (ii) The values of $H_{c2}(0)$ evaluated from $\mu$SR
and magnetization measurements coincide within the experimental
accuracy. (iii) Over the whole temperature range (from $T_c$ down to
30~mK) the temperature dependence of $\lambda$ is consistent with
what is expected for a single-gap s--wave BCS superconductor. (iv)
No influence of the applied magnetic field to $\lambda(T)$ was
observed.  (v) At $T=0$, the magnetic field dependence of
$\sigma_{sc}$ is in agreement with what is expected for a
superconductor with an isotropic energy gap.

To conclude, all the above mentioned features suggest that
Li$_2$Pd$_3$B is a {\it BCS superconductor with an isotropic energy
gap}.

\section{Acknowledgments}

This work was partly performed at the Swiss Muon Source (S$\mu$S),
Paul Scherrer Institute (PSI, Switzerland). The authors are grateful
to  S.~Str\"assle for help during manuscript preparation. This work
was supported by the Swiss National Science Foundation.

\end{document}